# Thermal rectification optimization in nanoporous Si using Monte Carlo simulations

Dhritiman Chakraborty[*], Joshua Brooke, Nick C S Hulse and Neophytos Neophytou

[1]School of Engineering, University of Warwick, Coventry, CV4 7AL, UK

[*] D.Chakraborty@warwick.ac.uk

## Abstract

We investigate thermal rectification in nanoporous silicon using a semi-classical Monte Carlo (MC) simulation method. We consider geometrically asymmetric nanoporous structures, and investigate the combined effects of porosity, inter-pore distance, and pore position relative to the device boundaries. Two basis geometries are considered, one in which the pores are arranged in rectangular arrays, and ones in which they form triangular arrangements. We show that systems: i) with denser, compressed pore arrangements (i.e with smaller inter-pore distances), ii) with pores positioned closer to the device edge/contact, and iii) with pores in a triangular arrangement, can achieve rectification of over 55%. Introducing smaller pores into existing porous geometries in a hierarchical fashion increases rectification even further to over 60%. Importantly, for the structures we simulate, we show that sharp rectifying junctions, separating regions of long from short phonon mean-free-paths are more beneficial for rectification than spreading the asymmetry throughout the material along the heat direction in a graded fashion.

# I. Introduction

Recently, significant research has been done on understanding and controlling phonon transport in nanostructures for novel materials and applications [1-8]. Since initial findings of thermal rectification between Cu and $Cu_2O$ [9] interfaces in the 1930s, various experimental [10-12] and theoretical [13-21] studies have investigated thermal rectification in different materials. Rectification values of up to 350% were theoretically predicted in graphene nanoribbons [22], while experiments showed that graphene junctions could provide even higher values of up to 800% rectification [12]. In the more technologically placed silicon, theoretical studies suggest that geometrically asymmetric structures can enhance thermal rectification effects [14-19], verified by some experimental works as well [10, 15, 23]. Rectification is achieved when the structure is separated into regions in which the mean-free-path (MFP) is controlled by two mechanisms – the temperature-dependent Umklapp scattering, and a mechanism much less temperature dependent such as boundary scattering. To-date, however, the specifics that determine rectification, as well as the design details which would allow further optimization, are not well understood. In this work, using Monte Carlo phonon transport simulations, we intend to provide the details and important elements that control thermal rectification in nanostructured materials.

# II. Approach

We use nanoporous Si as an example, and place the pores in various geometrical configurations to identify the parameters that determine rectification. The basis geometry we begin with is shown in Fig. 1. For computational efficiency we consider a two-dimensional (2D) simulation domain of length $L_x$ = 1000 nm in the *x*-direction and width $L_y$ = 500 nm in the *y*-direction. The asymmetry is created by placing the pores only in one part (porous region) of the material. We investigate the influence of porosity, pore density (as a function of inter-pore distance, $d$) and pore position relative to the device left/right boundary. Porosity ($\varphi$) is given as the total area of the pores (number of pores × area of each pore) over the total simulation domain area (given by the width × total length i.e. $L_T$ in Fig. 1).

We compute the thermal conductivity in the materials by solving the phonon Boltzmann transport equation using the Monte Carlo method as we described in our previous works [24-26]. The Monte Carlo (MC) approach offers great flexibility of geometrical configurations and parametric control over the scattering mechanisms that the phonons undergo, while still allowing very good accuracy [24, 26] and large micrometer size simulation domains. Further details on the Monte Carlo simulation approach we employ and the assumptions made therein can be found in the Appendix.

In the structures we consider, the domain is populated with circular pores of different arrangements. Each pore has a diameter of 50 nm. The choice of 50 nm is indirectly related to the phonon mean-free-path of pristine Si, $\lambda_{pp}$ = 135 nm, mostly though, for computational reasons: namely, for achieving diffusive transport we need channel dimensions of the order of several mean-free-paths, but not necessarily drastically larger to keep computational time manageable. Then, we need to deploy several pores into the domain, so we have used diameters smaller than the mean-free-path, but not so extremely small as to question the validity of the particle-based Monte Carlo method we employ. It seems that $D$ = 50 nm is a size to satisfy these criteria, and that is the size we employed in several of our works over the years as well [41, 43, Chakraborty18, Chakraborty19JEM, Chakraborty19MT].

All phonons reflect on the pores, with the pore boundaries taken to scatter all incident phonons in a specular fashion [24]. While systems with asymmetric roughness do have a thermal rectification effect [27], in this work we leave the effect of roughness aside, and focus on the geometrical configurations of the pores. The basis simulation domain with 8% porosity is shown in Fig. 1. The coloring indicates the thermal gradient between the hot side with temperature $T_H$ = 310 K (yellow) and the cold side with $T_C$ = 290 K (green). The green arrow on top of the schematic indicates the 'Forward' direction of heat flow, which we define when the pores are placed near the left, 'hot' contact, while keeping the 'cold' side empty of pores, thus creating an asymmetry in the transport direction. The thermal conductivity in this Forward direction is denoted as $\kappa_F$. Next, the structure is 'flipped' by rotating it 180 degrees such that the pores are now on 'cold' side while keeping the 'hot' side empty of pores. The thermal conductivity in this 'Reverse' direction is denoted as $\kappa_R$. The rectification, $R$, is defined as:

$$R = \frac{\kappa_F}{\kappa_R} - 1 \qquad (1)$$

We do not consider any modifications in the phononic bandstructure in the presence of pores, even if they are placed in periodic arrays. For the bandstructure to be modified, the requirement is that the wave-type carriers should remain coherent in distances several time larger than their wavelengths, as well as through several periods of the pore structure that forms the 'metamaterial'. Phonon-phonon scattering (with mean-free-path ~ 135 nm in Si) destroys coherence. The Monte Carlo approach we use is particle based and does not capture wave effects to begin with, however we are confident that ignoring bandstructure modifications is fully justified [28, 29].

## III. Results

<u>Rectangular nanoporous arrangements – influence of asymmetry:</u> We begin our investigation by considering rectangular arrangements of pores as in Fig. 1 and the schematics in Fig. 2. Figure 2 summarizes the effect of asymmetry on the thermal rectification ($R$) in this type of geometrical configuration. To change the asymmetry, we begin with placing pores on the left of the domain/material, and then gradually add more and more layers separated by an inter-pore distance $d$, until the material becomes fully porous symmetric. We denote this asymmetry as the increasing average porosity ($\varphi$) of the material with the addition of porous layers. Examples of the typical geometries considered, with porosities of 6% and 10% for both 'Forward' and 'Reverse' configurations, are shown in the geometry panels above Fig. 2. In Fig. 2 we plot the thermal conductivity $\kappa$ as a function of porosity $\varphi$ on the left $y$-axis. The 'Forward' direction thermal conductivity ($\kappa_F$) is plotted by the red line, whereas 'Reverse' $\kappa_R$ is by the blue line. The rectification $R$, is given by the purple line and indicated on the right $y$-axis.

First, we observe that the largest difference between $\kappa_F$, and $\kappa_R$ is observed for the lower porosity case, where the pores are placed all the way to the edge of one side. For that case ($\varphi = 6\%$) a difference of ~ 30 W/mK between $\kappa_F$ and $\kappa_R$ is observed. This gives the highest rectification value of 25% for the geometries in Fig. 2. As more porous columns are added and the overall porosity increases, the structure becomes more symmetric, the thermal conductivity decreases, but the difference in $\kappa_F$ and $\kappa_R$ also decreases, which reduces rectification. At $\varphi = 18\%$ the pores are uniformly distributed throughout the simulation domain and no rectification is observed. Thus, structural asymmetry in the transport direction (*x*-direction) gives rise to thermal rectification as also observed in previous theoretical [16, 17, 20] and experimental works [30, 31]. In Fig. 2, we can clearly observe that the $\kappa_F$ is larger compared to $\kappa_R$, which means that the hot phonons with shorter MFPs encountering the pores suffer less. In the $\kappa_R$ situation (when the structure is 'flipped'), the phonons that propagate from the pristine left (hot) to the porous right (cold), however, suffer more as by the time they reach the cold right side they develop longer MFPs.

This behaviour can be understood through a simple application of Matthiessen's rule. Under this assumption, we consider the fact that *R* is actually determined by the change in transport properties of phonons as they encounter the pores while at high temperature regions (shorter MFPs), or while at lower temperature regions (longer MFPs). A simple model that can predict this behaviour is obtained by splitting the material system into two regions of length $L_i$, that determine the total thermal resistance when added together as $\rho_T L_T = \rho_1 L_1 + \rho_2 L_2$, where $\rho_i$ are the thermal resistivities of each region which are weighted by the length of each region. Equivalently:

$$\frac{L_T}{\kappa_T} = \frac{L_{\text{PRISTINE}}}{\kappa_{\text{PRISTINE}}} + \frac{L_{\text{POROUS}}}{\kappa_{\text{POROUS}}} \quad (2)$$

The thermal conductivities in each region are proportional to the MFPs of phonons in each region as coupled together using Matthiessen's rule as:

$$\frac{1}{\kappa_T} \sim \frac{L_{\text{PRISTINE}}}{\lambda_{\text{PRISTINE}}} + \frac{L_{\text{POROUS}}}{\lambda_{\text{POROUS}}} \quad (3)$$

where $\lambda_{\text{PRISTINE}}$ is the temperature dependent Umklapp 3-phonon scattering MFP for Si ($\lambda_{\text{PRISTINE}} = \lambda_{pp} = 135$ nm at T = 300 K) and $\lambda_{\text{POROUS}}$ is the MFP in the porous region, which is given by the combination of two different scattering mechanisms, the Umklapp scattering MFP $\lambda_{pp}$ and the pore scattering MFP, which we take as *d*, the average distance between the pores. Thus, the MFP in the porous region is given by:

$$\frac{1}{\lambda_{\text{POROUS}}} = \frac{1}{\lambda_{pp}} + \frac{1}{d} \quad (4)$$

Considering the thermal conductivity in the Forward direction ($\kappa_F$), when pores are placed closer to the hot junction as in Fig. 1, we then have:

$$\kappa_F \sim \left(\frac{1}{\frac{1}{\lambda_H}+\frac{1}{d}}\right)\frac{L_H}{L_T} + \lambda_C \frac{L_C}{L_T} \quad (5)$$

where $\lambda_H < \lambda_{pp} < \lambda_C$ since the two regions are on the hotter/colder sides of the average temperature. Similarly, in the reverse direction we have:

$$\kappa_R \sim \lambda_H \frac{L_H}{L_T} + \left(\frac{1}{\frac{1}{\lambda_C}+\frac{1}{d}}\right)\frac{L_C}{L_T} \quad (6)$$

Inserting Eq. 5 and Eq. 6 in Eq. 1 we can estimate the rectification $R$.

The dashed-black line in Fig. 2 shows the rectification $R$ calculated using Eq. 1 with $\lambda_H$ = 117 nm for the hot side ($T_H$ = 310 K) as and $\lambda_C$ = 150 nm for the cold side ($T_C$ = 290 K), taken from Jeong *et* al. [32] and verified by our simulator. Note that the model only provides the rectification $R$, and not thermal conductivity values, as we only have proportionalities to the MFPs in Eq. 5 and Eq. 6. Using Eq. 1 and the lengths of the different porous regions, $R$ values for all porosities are calculated, and a very surprising match to the Monte Carlo results is observed (black versus purple lines in Fig. 2) using the values of $\lambda_H$ and $\lambda_C$ above.

Thus, based on this reasoning, what is important for rectification is the formation of the regions which force phonons from the hot side in the 'Forward' direction to interact as differently as possible with phonons reaching there from the cold side during the 'Reverse' direction. The closer the porous region is to the contacts, the larger the differences between the $\lambda_H$ and $\lambda_C$, which in our simulations varies almost linearly between the two contacts [32, 33], and the larger the rectification is (purple line in Fig. 2). Note that studies also suggest that a non-linear temperature gradient can lead to rectification [16, 21], in our case, however, it is the variation of the phonon MFPs that does so. Another important observation is the monotonic decrease in rectification as the pore columns pile up across the simulation domain, in the transport direction, in which case the structure becomes more and more 'symmetric'. The longer the porous region, however, the smaller the difference between $\lambda_H$ and $\lambda_C$ (if we consider at first order $\lambda_H$ and $\lambda_C$ as the mean-free-paths in the middle of the corresponding porous and the pristine regions, respectively), and rectification decreases.

Influence of pores on mean-free-paths: In order to further clarify the role that the phonon mean-free-paths (MFPs) play in rectification we look at the spectral propagation of phonons through the network of the pores. The overall MFP of phonons is extracted in two places in the material in the middle of the porous region and in the pristine material region, as indicated by the vertical lines in the insets of Fig. 3. We then extract the MFP distributions for the case of the 'Forward' and the 'Reverse' structure cases. These are shown in Fig. 3a and 3b, respectively. To do this, for every phonon that crosses the dashed vertical lines we record the absolute time that the phonon last scattered before crossing the line, and the time that the phonon scatters after crossing the line. Using this interval time and the phonon velocity, we compute the phonon MFP. Thus, we have four MFP distributions, two for the 'Forward' structure (Fig. 3a) and two for the 'Reverse' structure (Fig. 3b). We label the different cases '$F_H$' (for forward-hot porous side – solid red line in Fig. 3a), '$F_C$' (for forward-cold pristine side – solid blue line in Fig. 3a), '$R_H$' (for reverse-hot pristine side – dashed red line in Fig. 3b), and '$R_C$' (for reverse-cold porous side – dashed blue line in Fig. 3b). These distributions are the combined effective scattering MFP of the phonons due to phonon-phonon scattering and pore boundary scattering.

In the pristine non-porous regions the MFP is equivalent to the pristine $\lambda_{pp}$ of phonons, but in the porous regions the overall MFP $< \lambda_{pp}$ since pore boundary scattering is also added. When the device is 'flipped' from the 'Forward' to the 'Reverse' case, the spectrum of the 'hot' side shifts from the solid red line to the dashed-red line and for the 'cold' side it shifts from the solid blue line to the dashed-blue line. The asymmetry in the shift of cold-to-hot and hot-to-cold with respect to MFPs points directly to the reason that rectification is created. From the data in Fig. 3a, in the '$F_H$' case (porous region) the MFP is ~ 29 nm (compared to 33 nm from the analytical evaluation of Matthiessen's rule with $\lambda_H$ = 117 nm and $d$ = 50 nm). When the device is 'flipped' in Fig. 3b, the '$R_C$' case (porous region) has MFP ~ 33 nm (compared to 37.5 nm from the analytical evaluation of Matthiessen's rule with $\lambda_C$ = 150 and $d$ = 50 nm). Interestingly, the Monte Carlo computed MFP changes by 17% when the device is 'flipped', (i.e MFP$_F$/ MFP$_R$ - 1 = ~17%) very similar to the 18% rectification we compute for this structure.

Rectangular nanoporous arrangements - influence of pore positions and density: To better understand the geometric details that influence rectification further, we perform a series of simulations by altering the position, inter-pore spacing, and porous' regions effective resistance (the last is achieved by changing from an aligned to a staggered pore configuration) [26]. In Fig. 4 we show the rectification achieved by the porous geometries illustrated in the depicted schematics. The first series of simulations begin with the rectangular arrangement of pores located at the left side at a distance of 200 nm (~2 MFPs) from the left contact ($x = 200$ nm). The pores are located at distances $d = 50$ nm with respect to each other (light-blue, upper-left most schematic). We then investigate the effect of geometrical features on the rectifications as follows: i) by changing inter-pore distance, $d$, and making the structures more 'compressed', thus increasing thermal resistance; (still keeping the number of pores the same), ii) by staggering the overall geometry, thus reducing the line-of-side [34-36], also increasing the pore region resistance. iii) by combining effects (i) and (ii) i.e. reducing the inter-pore distance and staggering the pores.

First, we look at the rectification results shown by the light-blue bars ($x = 200$ nm) in Fig. 4. We start from the basis (rectangular) arrangement of pores (labelled arrangement 'A' on the figure) with a porosity of 8% and pore diameter $D = 50$ nm which provides 11% rectification. Reducing the inter-pore distance (arrangement 'B') increases rectification to 20%, then by staggering every alternate line of pores i.e. by moving the pores by 50 nm in the $y$-axis (arrangement 'C'), an additional small rectification is added to reach 24%. The combination of the two further increases rectification to 31% (arrangement 'D') (Note that in this last case we reduced the inter pore distance even further to $d = 12.5$ nm). This indicates that the larger the resistance of the rectifying region/junction, the larger the rectification that can be achieved.

Next, in the second (dark-blue, $x = 100$ nm) cases, we shift the entire pore geometry by 100 nm and place it at 100 nm from the domain edge (~1 MFP). We observe that just by moving the pores closer to the channel edge, the thermal rectification increases from the initial 11% to 18% (left-most structures, arrangement 'A'). This shift of the porous region brings a substantial increase, and we can explain it as follows: The placement of the pores in different regions in the channel, with different temperatures, interacts with phonons of different MFPs. In the dark-blue, $x = 100$ nm case, for example, in the 'Forward' direction the phonons encountering the pores have MFPs $\lambda_H$ = 120 nm (> $\lambda_{pp}$ = 135 nm since T = 308 K), while in the 'Reverse' direction, they have MFP $\lambda_C$ = 147 nm (since T = 292 K). On the other hand, in the light-blue, $x$ = 200 nm case, the numbers are slightly closer together, i.e. $\lambda_H$ = 124 nm (T = 306 K) in the 'Forward' direction and $\lambda_C$ = 143 nm (T = 294 K) in the reverse direction. (Note that here we have considered the temperature of the beginning of the pore region that is closer to the corresponding contact to report the MFPs). The larger the differences in the MFPs of the phonons when they encounter the pores in the 'Forward and 'Reverse' directions, the larger the rectification, and that is why the systems which place the pores closer to the device edge have larger rectification. The simple reason is that the longer the phonon MFPs, the more phonons are affected by nanostructuring (within the Matthiessen's rule picture), and the larger the difference between the MFPs of the phonons when they encounter the porous regions in the two configurations, the larger the rectification. As a simple first order verification, inserting the above values in the model of Eqs. 5-6, provides values for $R$ ~ 18% for dark-blue, $x$ = 100 nm case and $R$ ~ 13% for the light-blue, $x$ = 200 nm case, very similar to the ones that result out of the simulation ($R$ ~ 18% and $R$ ~ 11%, respectively). It seems that the choice of picking the MFPs at the beginning of the porous regions closer to the contacts provides this very good match. Following the same geometrical arrangements, 'A' – 'D' in the dark-blue cases (like in the light-blue, cases) we observe an increased rectification for the $x$ = 100 nm cases at all cases (compare the dark-blue bars to the light-blue bars). Bringing the pores closer together to $d/2$ = 25 nm in the $x$-direction ('B' in Fig. 4) increases rectification to 28%, introducing staggering ('C') provides another increment to 35%, and combining the two geometries (and enforcing an even smaller separation) as in the 'D' arrangement, the overall rectification jumps to as high as 44%.

In order to better understand the increase in rectification as we move pores closer to the device edge, we compare the contributions of $\kappa_F$ and $\kappa_R$ to rectification of the same structures in Fig. 4. In Fig. 5 we show the corresponding $\kappa_F$ and $\kappa_R$ by the red and blue bars, for the same structures "A" - "D" respectively. Figure 5 shows that, as expected, both $\kappa_F$ and $\kappa_R$ drop as the structure becomes more resistive, but the reduction in $\kappa_R$ (blue bars) is larger for all cases, and this is the primary reason behind the rectification improvements. For instance, the $\kappa_R$ observed in the light-blue, $x$ = 200 nm case drops by ~29% when comparing the basis left-most structure ('A') to the right-most one ('D'). On the other hand, the corresponding $\kappa_R$ drop in the dark-blue, $x = 100$ nm case is ~36%. We can draw a couple of conclusions from these observations: i) the larger drop in $\kappa_R$ compared to that of $\kappa_F$ is what increases rectification in these structures, and ii) that $\kappa_R$ reduces more as pores are brought closer to the cold domain edge. That is again in agreement with the simple Matthiessen's rule based consideration, which states that it is the long MFP colder phonons that are affected more when encountering the porous regions.

Thus, the greatest rectification is seen for the most asymmetric case, where the temperature-dependent MFPs of phonons differ as much as possible between the 'Forward' and 'Reverse' directions. Hence, as placing the pores closer to the edge enhances rectification, from here on we consider the $x = 100$ nm basis case to examine further geometrical configurations.

Influence of neck/diameter geometrical considerations and temperature: To further understand the interplay between the geometrical features and the mean-free-path (MFP) in determining rectification, we examine the combination of pore diameter ($D$) and the corresponding interpore distance (neck, $n$) in determining thermal resistance in these nanostructures (our geometries so far have $n = 50$ nm and $D = 50$ nm, which corresponds to to ratio $n/D = 1$). Different diameters and neck sizes will interact with different phonon MFPs differently. In Fig. 6 we show the rectification simulation results for structures with different pore sizes (and neck sizes) as well versus different $n/D$ ratios, for 3 different temperatures. Some typical geometries are indicated in the insets. First we note that that reduced $n/D$ ratio gives higher thermal resistance, which gives greater rectification until a 'cut-off' neck size.

The saturation in rectification for small $n/D$ is explained as follows: Rectification is achieved when the overall mean-free-path (MFP) near the hot side of the 'Forward' direction structure, is different than the MFP near the cold side of the 'reverse' direction structure. When the scattering mean-free-path (determined by the neck distance and diameter) is of the order of the phonon mean-free-path, $\lambda_{pp}$, then variations in the geometry will have an influence in the overall MFP. As the neck size is decreased to sizes much smaller than $\lambda_{pp}$, then boundary scattering dominates the overall MFP in either the 'Reverse' or the 'Forward' structures, the $\lambda_{pp}$ has little effect, and rectification saturates. Thus, reducing neck size below this saturation point further, gives diminishing returns for rectification since it is already much smaller than $\lambda_{pp}$ at that point.

In Fig. 6 we also show the effect of temperature. We have performed simulations for the same set of different geometries at two more temperatures, one higher ($T_H = 350$ K), and one lower ($T_H = 250$ K) than the nominal room temperature. For a given geometry, higher temperatures reduce the phonon $\lambda_{pp}$ due to stronger phonon-phonon scattering ($\lambda_{pp} = 100$ nm for $T_H = 350$ K), and increase their influence in determining thermal resistance over the pore scattering, which reduces rectification. In contrast, lower temperatures increase the phonon MFPs ($\lambda_{pp} = 220$ nm for $T_H = 250$ K), which reduce their influence in determining thermal resistance over the pore scattering (or the pore influence becomes bigger), which increases rectification.

A different material, other than Si, will have different phonon MFPs, and the geometry will then need to be adjusted accordingly, such that the dominant heat carrying phonons actually interact effectively with the underlying nanopore geometry. For example, in materials with short phonon MFPs (as also in the case of Si at high temperatures), the neck size needs to be smaller compared to materials with larger phonon MFPs in order to achieve the same level of rectification (similar to how Si with smaller phonon MFPs at elevated temperatures requires a smaller *n/D* ratio to achieve the same rectification compared to Si at lower temperatures with larger MFPs in Fig. 6). Thus, we believe our conclusions can be generally applied to other materials as well, in which case the underlying geometry dimensions need to be adjusted to the MFP of the specific material.

Influence of exposed surface area and grading: From the simple model presented in Eqs. 5-6, rectification is determined by the asymmetry in the different interaction of the hot/cold MFPs in the porous regions, which of course will have a certain length. An interesting point to examine here, is the interplay between the effect of increasing asymmetry along the transport direction in a gradual (graded) way, versus having sharp junctions with a given interface length to separate the porous and pristine regions (length in our case of 2D simulations – in 3D it would be interface area). In the first (graded) case the rectification region will increase throughout the material, but its local influence will be smaller, whereas in the second case, the rectification region is small, essentially rectification is dominated by the junction between the porous/pristine regions, but its strength is larger.

In the simulation results shown in Fig. 7 we begin again with the basis 'A' structure of 8% porosity. In order to examine the effect of increasing the interface area that separates the two regions, we rearrange the pores in an oblique configuration ('E') seen in the panel above Fig. 7. We find that increasing the interface area, i.e. increasing the surface area where rectification happens, increases $R$ to ~ 22% overall (comparing the 'A' with 'E' case bars in Fig. 7). By following the same logic for staggered pores ('C') and exposing both sides we consider a triangular arrangement of pores (arrangement 'F') which gives an amplified $R$ = 45% (as compared to $R$ = 34% of the simple staggered arrangement 'C').

However, by extending this structure to get a graded geometry throughout the domain (graded-triangular) arrangement 'G' in Fig. 7), does not lead to any more rectification improvements. In fact, $R$ drops to ~31% from 45% that is achieved by the triangular structure (case 'F'), which is more asymmetric and divide the domain into two discrete porous and non-porous segments. In the fully graded 'G' case phonons interact with the pores throughout the domain. It could be argued that that a graded porous structure would provide a small degree of rectification locally, but that would extend to the entire channel, and the aggregated effect could be significant. Our simulations show, however, the reverse, namely that that discontinuous regions (junctions) provide a greater rectification than graded geometries in our Si structures, as also indicated in other materials [12, 13, 17, 20, 37-40], as the differences between the phonon MFPs in the 'Reverse' and 'Forward' directions in each segment of the structure are now smaller.

Triangular nanoporous arrangements - influence of hierarchical nanopores: Here onwards we use the triangular geometry configurations with pores placed $x$ = 100 nm from the edge (green bars and green geometries in Fig. 8) which provide the maximum rectification, and compare them to our basis rectangular structures (dark-blue bars and dark-blue geometries in Fig. 8). The rectification for the basis (rectangular) structure 'A' is 18% and this increases to 45% for the triangular case, as we have already seen. We then explore the effects of: i) decreasing inter-pore spacing as before, and ii) introducing smaller pores in a hierarchical fashion to the previously considered geometries. In both cases we aim to increase the resistance of the porous region, a strategy that increases rectification, as we have observed earlier. The structures can be seen in the geometry panel above Fig. 8 (dark-blue for the rectangular and green for the triangular cases, respectively).

In the second structure column ('H') in Fig. 8, by introducing smaller nanopores of diameter $D = 10$ nm to the existing base structures in a hierarchical fashion we increase the effective pore density. This gives $R = 51\%$ for the triangular case compared to 24% in the rectangular case (green versus blue bars for structures 'H' of Fig. 8). We then reduce the inter-pore distance from $d = 50$ nm to $d/2 = 25$ nm in the *x*-direction (column 'C' in Fig. 8), thus making the structures more 'compressed' with increased thermal resistance. Structure 'C' in the triangular arrangement causes an increase to $R = 57\%$, compared to 28% in the rectangular 'C' case.

Finally, by combining the effects of density and hierarchical nanopores we get a maximum rectification of $R = 61\%$ in the combined hierarchical structure (arrangement 'I') for the triangular case in Fig. 8. Compared to the arrangement 'D', the compressed rectangular case given by the white bar with blue-dashed outline in Fig. 8, the triangular configuration provides a 35% increase in heat rectification. This further stresses the importance of the larger rectifying junction length/area, as well as the high thermal resistance of the porous region, rather than the graded porous configurations in the structures we have investigated.

## IV. Conclusions

In this work we examined the effect of pore density, position, junction surface area, and hierarchical nanostructuring on thermal rectification in nanoporous Si using Monte Carlo simulations. Rectification requires asymmetry in geometry. We show that rectification is optimized: i) when the temperature-dependent phonon-phonon scattering limited MFPs of phonons encountering the porous regions are as different as possible in the 'Forward' compared to the 'Reverse' directions, ii) when the porous regions are as resistive as possible, and iii) when the rectifying junction that separates the porous and pristine regions is as large as possible. Practically, we have shown that these conditions are achieved: i) by placing the porous region as near to the contact as possible, ii) by making the porous region as resistive as possible (denser, compressed, staggered, hierarchically placed pores), and iii) by using triangular rather than rectangular pore region configurations. By combining these effects, we showed that rectification values of greater than 60 % can be reached.

# V. Acknowledgements

This work has received funding from the European Research Council (ERC) under the European Union's Horizon 2020 Research and Innovation Programme (Grant Agreement No. 678763).

# VI. Appendix: Single Phonon Monte Carlo Method

This section gives details about the Monte Carlo approach we employ, its validation, and the assumptions related the scattering term used. Full details can be found in our previous works [24, 41] . The simulation process is as follows: The simulation domain of size $L_x$ = 1000 nm × $L_y$ = 500 nm is initialized with a 'Hot' and 'Cold' end, set at $T_H$ = 310 K and $T_C$ = 290 K respectively.

Phonons are then allowed to enter the simulation domain from either side and alternate between free flight and scattering. The time a phonon spends in the simulation domain is recorded as its 'Time-Of-Flight' (TOF). Phonons are initialized based on energy, frequency, polarization, and velocity. We use the dispersion relation $\omega(q)$ and corresponding group velocities $v_g(q)$ as described by Pop *et al.* [42] in Eq. A1 and Eq. A2 below:

$$\omega(q) = v_s q + c q^2 \tag{A1}$$

$$v_g = \frac{d\omega}{dq} \tag{A2}$$

where $q$ is the wave vector norm and $v_s$ and c are fitting parameters to match the thermal conductivity of bulk Si in the [100] direction. The dispersion coefficients we use are $v_s$ = 9.01 × 10$^3$ ms$^{-1}$ and c = -2 × 10$^{-7}$ m$^2$s$^{-1}$ for the longitudinal acoustic (LA) branch, and $v_s$ = 5.23 × 10$^3$ ms$^{-1}$ and c = -2.26 × 10$^{-7}$ m$^2$s$^{-1}$ for the transverse acoustic (TA) branches [43]. Following common practice, the contribution of optical phonons is neglected as they have low group velocities and do not contribute significantly to phonon transport [44-47].

Phonons in the simulation domain either scatter, or are in free flight. During free flight, the position *r* at time *t* of the phonon is given by the equation:

$$r(t_i) = r(t_{i-1}) + v_g \Delta t \tag{A3}$$

Scattering of phonons is caused either by interaction with geometrical features, or by three-phonon internal scattering (Umklapp processes). The three-phonon scattering, which is responsible for the change in the temperature of the domain, is computed in the relaxation time approximation and is a function of temperature and frequency, as [42-45, 49]:

$$\tau_{TA,\,U}^{-1} = \begin{cases} 0 & for\ \omega < \omega_{1/2} \\ \dfrac{B_U^{TA}\omega^2}{\sinh\left(\dfrac{\hbar\omega}{k_B T}\right)} & for\ \omega > \omega_{1/2} \end{cases} \qquad (A4)$$

where $\omega$ is the frequency, $T$ the temperature, $B_U^{TA} = 5.5 \times 10^{-18}$ s, and $\omega_{1/2}$ is the frequency corresponding to $q = q_{max}/2$. These equations and parameters are well-established and often used to describe relaxation time in phonon Monte Carlo simulations for Si [42-45, 48]. Three phonon scattering causes a change in the energy, and thus the temperature ($T$) of the simulation domain 'cell' where scattering took place (we use a 0.1 nm domain discretization). Every time this happens the 'cell' temperature either rises or falls. The link between energy and temperature is given by:

$$E = \frac{V}{W}\sum_p\sum_i \left(\frac{\hbar\omega_i}{\exp\left(\dfrac{\hbar\omega_i}{k_i T}\right)-1}\right) g_i D(\omega_i, p)\Delta\omega \qquad (A5)$$

where $\omega$ is the frequency, $T$ the temperature, $D$ the density of states at given frequency and branch polarization, and $g_p$ the polarization branch degeneracy, and V is the volume of the 'cell'. The dissipation/absorption of energy from each 'cell' in this way establishes a temperature gradient under a continuous flow of phonons (shown in Fig. 1 with the yellow to green color scheme). A scaling factor (W = $4\times10^5$) is also introduced to scale the number of phonons simulated to the real population of phonons from $6\times10^{10}$ μm$^{-3}$ that are present at 300 K for computational efficiency [41].

The total energy entering and leaving the simulation domain is calculated by the net sum of the corresponding phonon energies that enter/exit at the hot and cold junctions as calculated by Eq. A5. We label the total incident energy from the hot junction as $E_{in}^H$ and the total energy of phonons leaving the simulation domain from the hot junction as $E_{out}^H$. Similarly, $E_{in}^C$ and $E_{out}^C$ are the in-coming and out-going energies at the cold junction. We then determine the average phonon energy flux in the system as:

$$\Phi = \frac{\left(E_{\text{in}}^{\text{H}} - E_{\text{out}}^{\text{H}}\right) - \left(E_{\text{in}}^{\text{C}} - E_{\text{out}}^{\text{C}}\right)}{n\langle TOF \rangle} \tag{A6}$$

where $n$ is the total number of phonons simulated (typically 1 million) and $\langle TOF \rangle$ is the average time-of-flight of all phonons. The thermal conductivity ($\kappa$), is then calculated using Fourier's law.

$$\Phi = -\kappa_{\text{s}} \nabla T \tag{A7}$$

Next, to account for the fact that the length of the simulated domain ($L_x$) is smaller than some phonon wavelengths, especially at lower temperatures, a scaling of the simulated thermal conductivity ($\kappa_s$) is needed to compute the final thermal conductivity $\kappa$ as [**32**]:

$$\kappa = \kappa_{\text{s}} \frac{\left(L_{\text{X}} + \lambda_{\text{pp}}\right)}{\left(L_{\text{X}}\right)} \tag{A8}$$

where $\lambda_{\text{pp}}$ is the average phonon mean-free-path (MFP) of Si. In previous work simulations were carried out to compare and validate the simulator for bulk values of silicon thermal conductivity [24, 26, 41, 43]. In particular we have also shown that our simulator results compare very well with several literature simulation and experimental results for the thermal conductivity versus temperature for pristine bulk Si and porous Si cases [25, 26].

Figure 1:

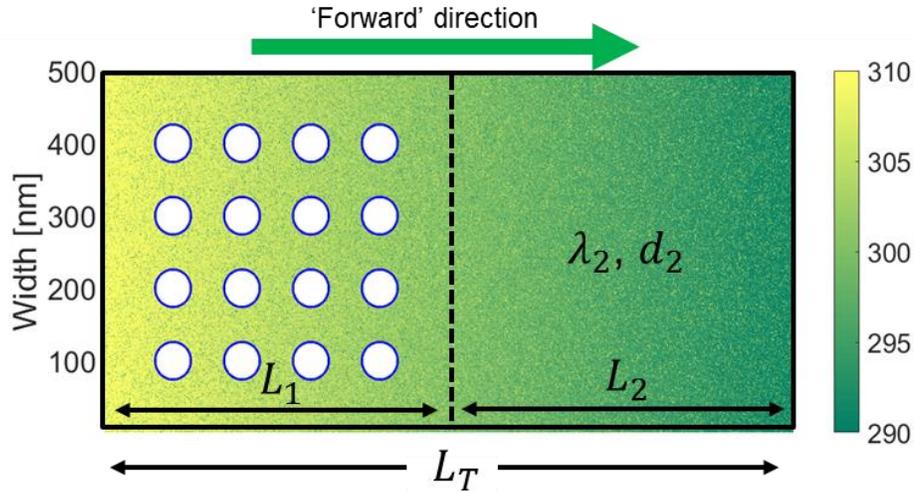

Figure 1 caption:

Schematic of the basis geometry simulated with porous region of length $L_1$ and pristine region of length $L_2$. Individual properties are assigned for each region, including average phonon mean-free-path, (MFP) $\lambda$ and average distance between pores (inter-pore distance) $d$. The total length of simulation domain is $L_T$. In all Monte Carlo simulations we set $L_T = 1000$ nm. The coloring indicates the established thermal gradients when the left and right contacts are set to $T_H = 310$ K (**yellow**) and $T_C = 290$ K (**green**), respectively. The **green arrow** above the schematic depicts "**Forward**" direction of heat flow from $T_H$ to $T_C$.

Figure 2:

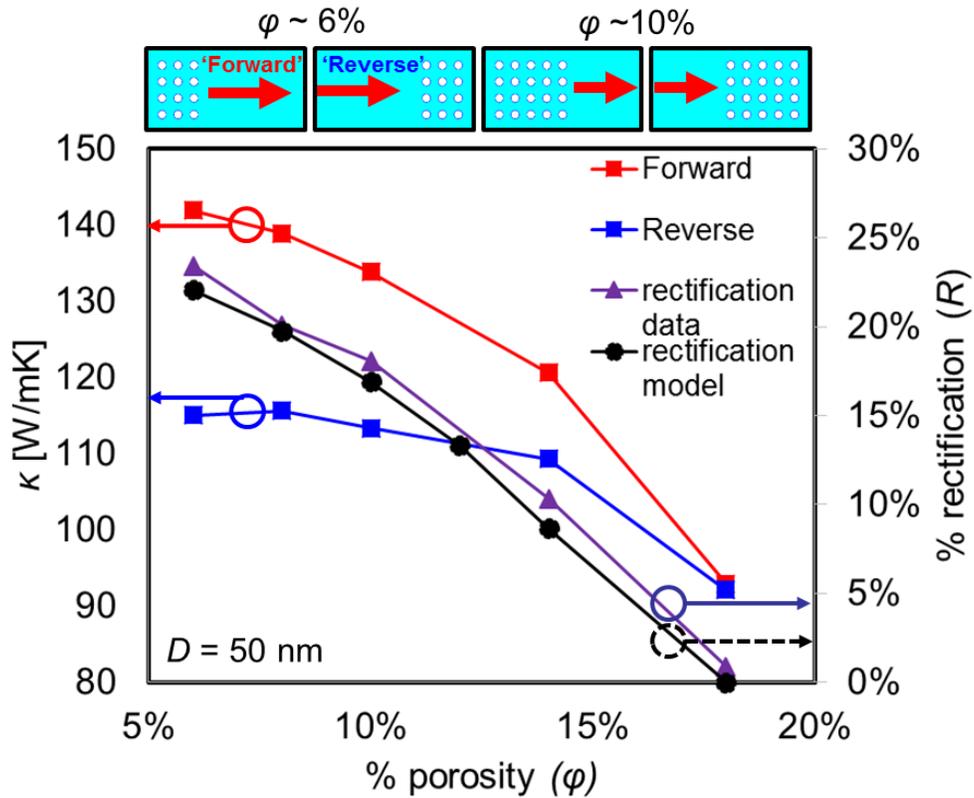

Figure 2 caption:

Monte Carlo simulations showing the effect of porosity ($\varphi$) in the rectangular, ordered pores configuration on thermal conductivity, $\kappa$ (left axis), and rectification, $R$ (right axis). For each value of $\varphi$, the $\kappa$ in the 'Forward' direction ($\kappa_F$) is given by the red line, while the $\kappa$ in the 'Reverse' direction ($\kappa_R$) is given by the blue line and the rectification data is given by the purple line. The dashed-black line gives the results predicted by the model given by Eqs. 1, 5-6. Examples of typical geometries simulated for 6% and 10% porosity are shown above the figure, as well as our definitions for 'Forward' and 'Reverse' directions. All pore diameters are 50 nm.

Figure 3:

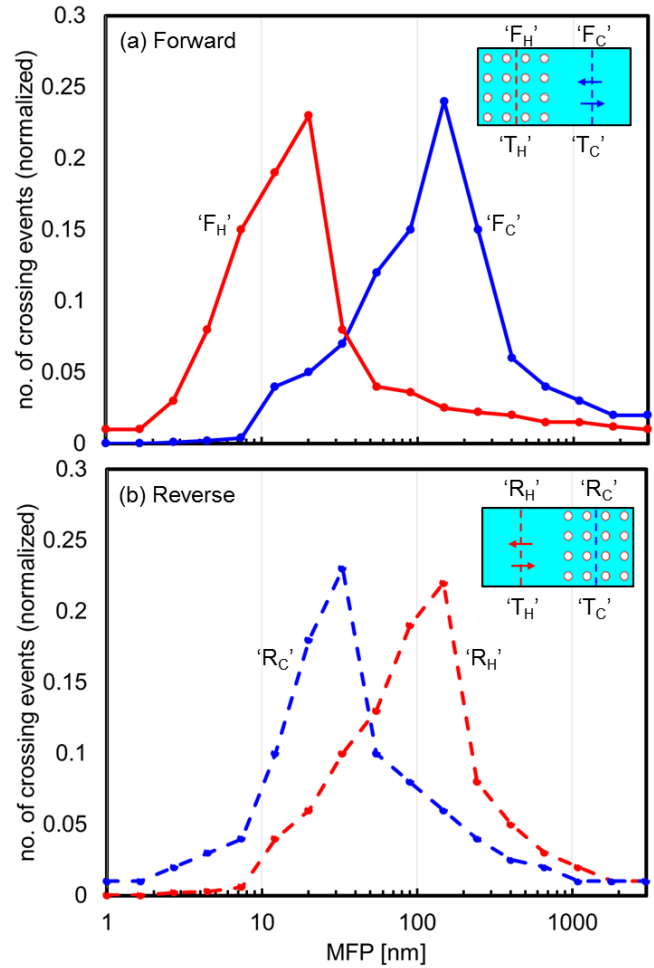

Figure 3 caption:

The MFP distribution for porous and non-porous regions. The geometry considered is given in the inset. The y-axis indicates normalized number of crossing events by phonons across the dashed lines (a) The 'Forward' structure case. (b) The 'Reverse' structure case red lines indicate the phonon MFP distribution near the hot sides, and blue lines near the cold sides, following the same coloring scheme as the vertical crossing lines in the insets.

Figure 4:

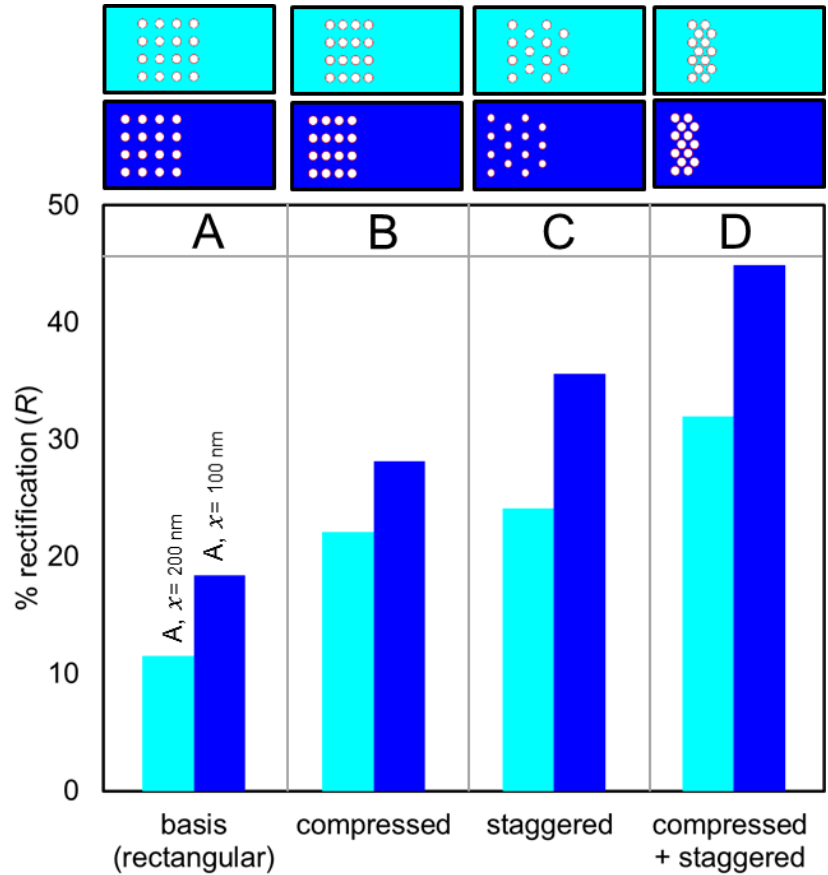

Figure 4 caption:

Monte Carlo simulations showing the effect of pore position, inter-pore distance compression and pore staggering on rectification ($R$ on left axis). Four cases are examined, and the geometries simulated are given in the panel above the figure. These are i) the basis (rectangular) arrangement of pores ('A') with pore diameter $D = 50$ nm; ii) compressed arrangement ('B') which has the same configuration as 'A', but with halved inter-pore separation; iii) staggered arrangement ('C') given by shifting pore positions of 'A' by 50 nm in the $y$-direction; iv) compressed + staggered arrangement of pores ('D') by reducing inter-pore distance to 12.5 nm in (iii). Pores are placed at 200 nm from the domain edge in the first (light-blue, $x = 200$ nm) cases in all four arrangements 'A' – 'D'. They are left-shifted by 100 nm and placed 100 nm from the domain edge in the second (dark-blue, $x = 100$ nm) cases in all four arrangements.

Figure 5:

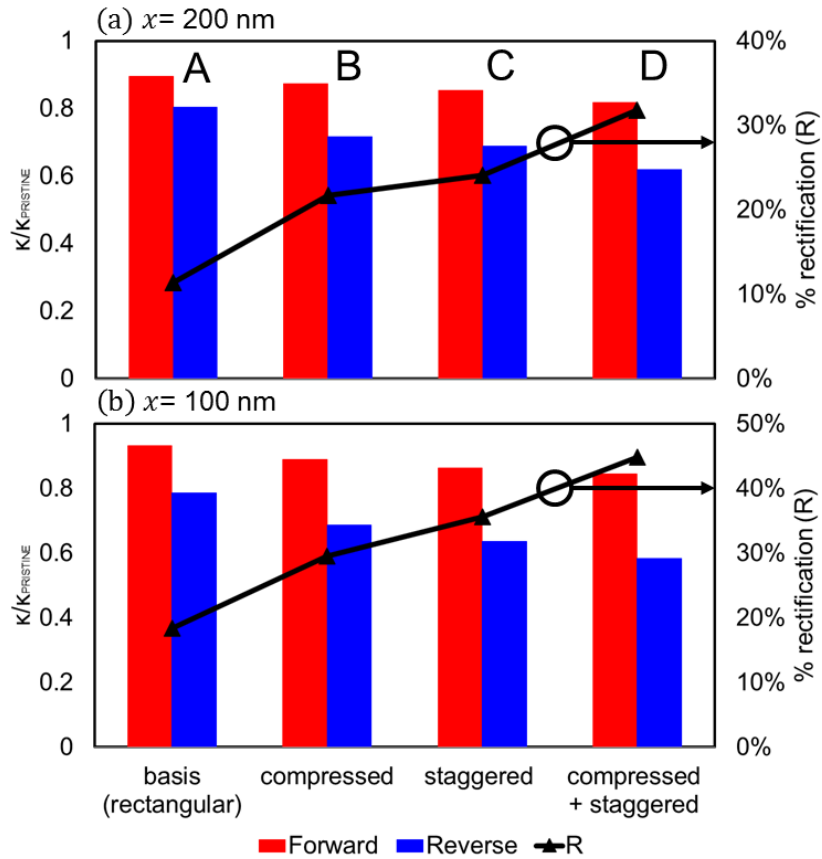

Figure 5 caption:

Normalized $\kappa$ observed in the 'Forward' (red bars) and the 'Reverse' (blue bars) direction for each structures 'A'-'D' of Fig. 4 (left axis). The rectification is shown by the black line in the right axis. (a) The pores are placed at 200 nm from the domain edge ($x = 200$ nm). (b) The pores are shifted by 100 nm and placed 100 nm from the domain edge ($x = 100$ nm). In each graph the $\kappa$ is normalized to $\kappa_{\text{PRISTINE}}$ (i.e. $\kappa = 148$ W/mK).

Figure 6:

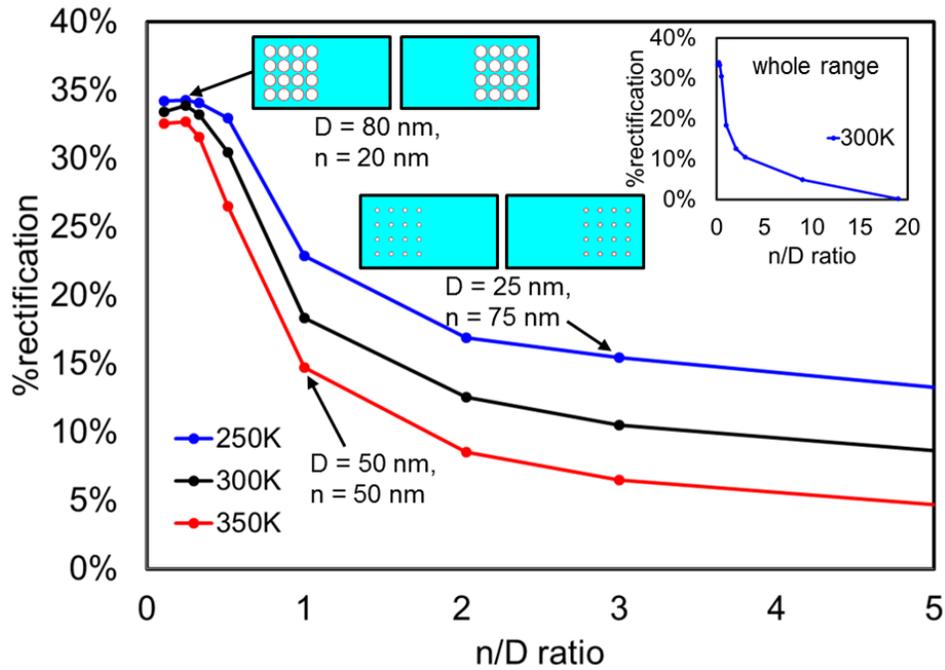

Figure 6 caption:

The rectification dependence (*y*-axis) on the pore and neck sizes – given by the neck/diameter (*n/D*) ratio (*x*-axis), and the mean temperature of the simulation domain. Three temperature ranges are examined $T_{avg}$ = 350 K (light-blue line) for $T_H$ = 360 K to $T_C$ = 340 K, $T_{avg}$ = 300 K (blue line) with $T_H$ = 310 K to $T_C$ = 290 K and $T_{avg}$ = 250 K (purple line) with $T_H$ = 260 K to $T_C$ = 240 K. The rectification for whole range of *n/D* values simulated for the $T_{avg}$ = 300 K (blue line) case is given as an inset on the top right corner. Typical geometries simulated are also indicated in the insets.

Figure 7:

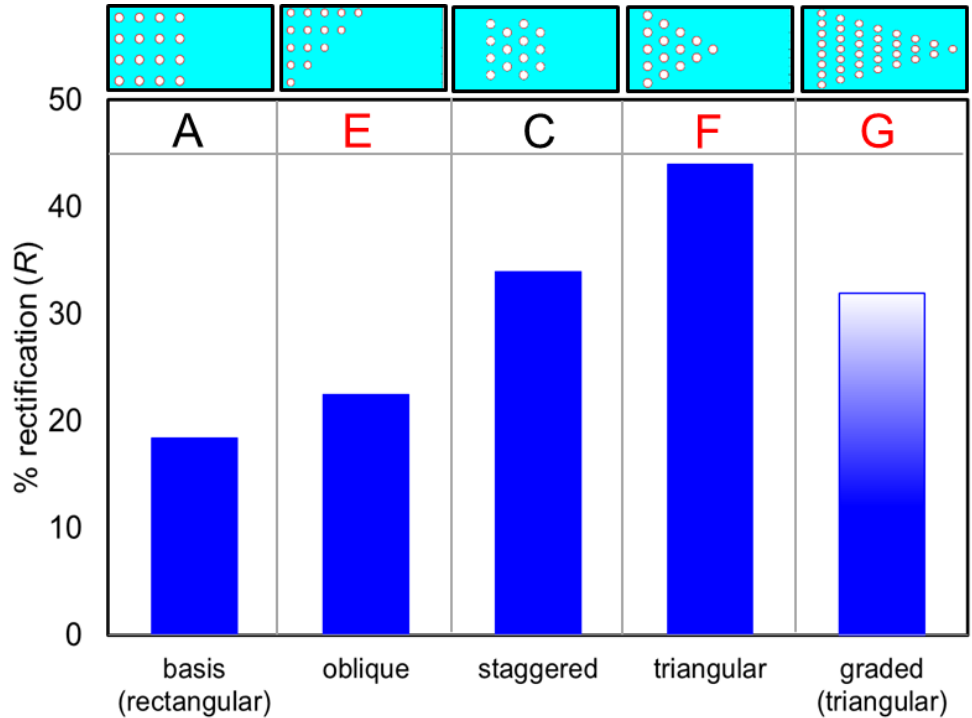

Figure 7 caption:

The dependence of thermal rectification on the exposed junction surface area and pore grading. Five cases are examined, and the geometries simulated are given above the bar chart. These are: i) the basis (rectangular) arrangement of pores ('A') with pore diameter $D$ = 50 nm; ii) oblique arrangement ('E') where pores are arranged in a right-triangular fashion to give increased exposed surface area; iii) staggered arrangement ('C'); iv) triangular arrangement ('F'); v) graded (triangular) arrangement ('G') of pores by uniformly decreasing pore density in the $x$ direction.

Figure 8:

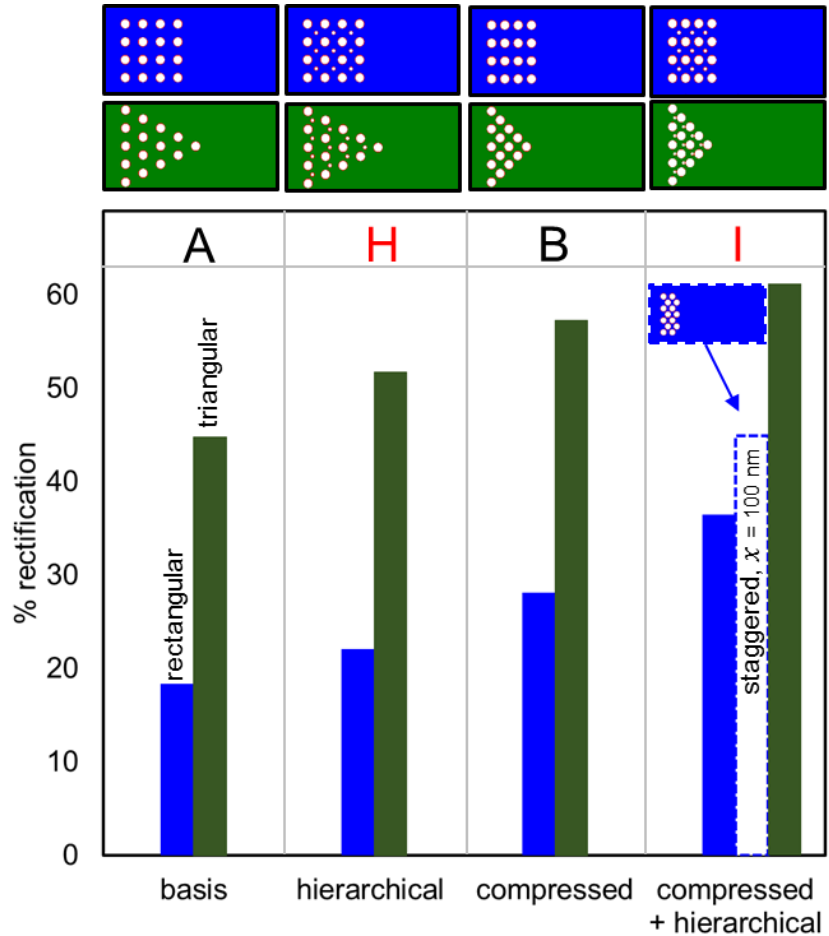

Figure 8 caption:

The rectification dependence on the construction of the triangular regions, and the hierarchical incorporation of smaller nanopores ($D = 10$ nm) in between the bigger ones ($D = 50$ nm). Four cases are examined, each for both rectangular geometries (dark-blue bars) and triangular geometries (green bars). The geometries simulated are shown in the panel above the figure. These are: i) the basis arrangement of pores ('A') for both rectangular and triangular configurations given by dark-blue bars or green bars, respectively. For all pores $D = 50$ nm, and inter pore distance $d = 50$ nm; ii) hierarchical arrangement ('H') by adding smaller pores in-between 'A'; iii) compressed arrangement ('B') with halved inter-pore separation $d = 25$ nm; iv) compressed + hierarchical arrangement of pores ('I') by halving the original inter-pore distance. The result for arrangement 'C' from Fig. 4 is also included for comparison. All structures are placed $x = 100$ nm from the edge of the device.